 \documentclass[final,1p,times]{elsarticle}

\usepackage{amssymb}
\usepackage{psfig}
\usepackage{amsmath}
\usepackage{amssymb}
\usepackage{graphicx}
\usepackage{subfigure}

\newcommand{\lSect}[1]{{\label{sec:#1}}}

\newcommand{\Sectff}[1]{{\ref{sec:#1}}}
\newcommand{\Sect}[1]{{\S~\Sectff{#1}}}

\newcommand{\note}[1]{\emph{\textcolor{red}{}}}

\newcommand{\Msun}{{\ensuremath{\mathrm{M}_{\odot}}}}

\newcommand{\Ni}{{\ensuremath{^{56}\mathrm{Ni}}}}

\newcommand{\Ep}[1]{{\ensuremath{10^{#1}}}}

\newcommand{\cm}{{\ensuremath{\mathrm{cm}}}}
\newcommand{\km}{{\ensuremath{\mathrm{km}}}}
\newcommand{\erg}{{\ensuremath{\mathrm{erg}}}}

\newcommand{\CASTRO}{\texttt{CASTRO}}
\newcommand{\FLASH}{\texttt{FLASH}}
\newcommand{\KEPLER}{\texttt{KEPLER}}
\newcommand{\MESA}{\texttt{MESA}}
\newcommand{\K}{\ensuremath{\mathrm{K}}}
\newcommand{\gcc}{\ensuremath{\mathrm{g}\,\mathrm{cm}^{-3}}}

\journal{Astronomy and Computing}

\begin{document}

\begin{frontmatter}



\title{Numerical Approaches for Multidimensional Simulations of Stellar Explosions}


\author [ucsc,umn]{Ke-Jung Chen\corref{cor1}}
\author [moca]{Alexander Heger}
\author [lbl]{Ann S. Almgren }

\cortext[cor1]{IAU Gruber Fellow, corresponding author; kchen@ucolick.org}
\address[ucsc]{Department of Astronomy \& Astrophysics, University of California, Santa Cruz, CA 95064, United States}
\address[umn]{School of Physics and Astronomy, University of
  Minnesota, Minneapolis, MN 55455,  United States}
\address[moca]{Monash Centre for Astrophysics, Monash University, Victoria 3800, Australia}
\address[lbl]{Computational Research Division, Lawrence Berkeley National
  Lab, Berkeley, CA 94720,  United States}

\address{}

\begin{abstract}
We introduce numerical algorithms for initializing multidimensional simulations of stellar explosions 
with 1D stellar evolution models.  The initial mapping from 1D profiles onto multidimensional grids can 
generate severe numerical artifacts, one of the most severe of which is the violation of conservation 
laws for physical quantities.  We introduce a numerical scheme for mapping 1D spherically-symmetric 
data onto multidimensional meshes so that these physical quantities are conserved.  We verify our 
scheme by porting a realistic 1D Lagrangian stellar profile to the new multidimensional Eulerian hydro 
code \CASTRO.  Our results show that all important features in the profiles are reproduced on the new 
grid and that conservation laws are enforced at all resolutions after mapping.  We also introduce a 
numerical scheme for initializing multidimensional supernova simulations with realistic perturbations 
predicted by 1D stellar evolution models.  Instead of seeding 3D stellar profiles with random 
perturbations, we imprint them with velocity perturbations that reproduce the Kolmogorov energy 
spectrum expected for highly turbulent convective regions in stars.  Our models return Kolmogorov 
energy spectra and vortex structures like those in turbulent flows before the modes become nonlinear.  
Finally, we describe approaches to determining the resolution for simulations required to capture fluid 
instabilities and nuclear burning.  Our algorithms are applicable to multidimensional simulations 
besides stellar explosions that range from astrophysics to cosmology.  

\end{abstract}


\end{frontmatter}
\section{Introduction}

Multidimensional simulations shed light on how fluid instabilities arising in supernovae (SNe) 
mix ejecta \citep{herant1994,candace2009,candace2010,candace2011}.  Unfortunately, computing 
fully self-consistent 3D stellar evolution models, from their formation to collapse, for the 
explosion setup is still beyond the realm of 
contemporary computational power.  One alternative is to first evolve the main sequence star in a 
1D stellar evolution code in which the equations of momentum, energy and mass are solved on a 
spherically-symmetric grid, such as \KEPLER{} \citep{kepler} or \MESA{} \citep{mesa}.  Once the 
star reaches the pre-supernova phase, its 1D profiles can then be mapped into multidimensional 
hydro codes such as \CASTRO{} \citep{ann2010,zhang2011} or \FLASH{} \citep{flash} and 
continue to be evolved until the star explodes.

Differences between codes in dimensionality and coordinate mesh can lead to numerical issues 
such as violation of conservation of mass and energy when profiles are mapped from one code 
to another.  A first, simple approach could be to initialize multidimensional grids by linear 
interpolation from corresponding mesh points on the 1D profiles.  However, linear interpolation 
becomes invalid when the new grid fails to resolve critical features in the original profile such as 
the inner core of a star.  This is especially true when porting profiles from 1D Lagrangian codes, 
which can easily resolve very small spatial features in mass coordinate, to a fixed or adaptive 
Eulerian grid.  In addition to conservation laws, some physical processes such as nuclear burning are 
very sensitive to temperature, so errors in mapping can lead to very different outcomes for 
the simulations such as altering the nucleosynthesis and energetics of SNe.  None address the conservation 
of physical quantities by such procedures.  We examine these issues and introduce a new scheme for 
mapping 1D data sets to multidimensional grids.  

Seeding the pre-supernova profile of the star with realistic perturbations is important to 
illuminate how fluid instabilities later erupt and mix the star during the explosion.  Massive 
stars usually develop convective zones prior to exploding as SNe \citep{woosley2002,heger2002}.  
Multidimensional stellar evolution models suggest that the fluid inside the convective regions can 
be highly turbulent \citep{porter2000,arnett2011}.  However, in lieu of the 3D stellar evolution 
calculations necessary to produce such perturbations from first principles, multidimensional 
simulations are usually just seeded with random perturbations.  In reality, if the star is convective 
and the fluid in those zones is turbulent \citep{davidson2004}, a better approach is to imprint the 
multidimensional profiles with velocity perturbations with a Kolmogorov energy spectrum \citep{frisch1995}.

In addition to implementing realistic initial conditions, care must be taken to determine the resolution  that
multidimensional simulations require to resolve the most important physical scales and yield 
consistent results given the computational resources that are available.  We provide a systematic 
approach for finding this resolution for multidimensional stellar explosions.  The structure of the paper is
as follows; in \Sect{model} we describe the key features of the \KEPLER{} and \CASTRO{} codes.  We describe 
our initial mapping scheme and demonstrate it by porting a massive star model from \KEPLER{} to \CASTRO{} 
in \Sect{mapping}.  We review our scheme for seeding 2D and 3D stellar profiles with turbulent perturbations and 
present hydrodynamic simulations done with these profiles in \CASTRO{} in \Sect{perb}.  
We provide a strategy for finding the proper resolution for multidimensional simulations with 
multiscale processes such as hydrodynamics and nuclear burning in \Sect{resolution} and conclude the results 
in  \Sect{conclusions}.  

\section{Stellar Model}
\lSect{model}

We model the evolution of main sequence stars with \KEPLER{} \citep{kepler}, a 1D Lagrangian 
stellar evolution code.  \KEPLER{} solves the evolution equations for mass, momentum, and 
energy, including relevant physical processes such as nuclear burning and mixing due to convection.  When 
the star reaches the pre-supernova phase (hundreds of seconds prior to launching the SN shock), 
we map its 1D profiles onto a multidimensional grid in \CASTRO{}.  When the star explodes, its initial 
spherical symmetry is broken by fluid instabilities formed during the explosion that cannot be 
modeled by 1D calculations.  Hence, we follow the evolution of the star in \CASTRO{} until it explodes.  

\begin{figure}[h]
\begin{center} 
\includegraphics[width=\columnwidth]{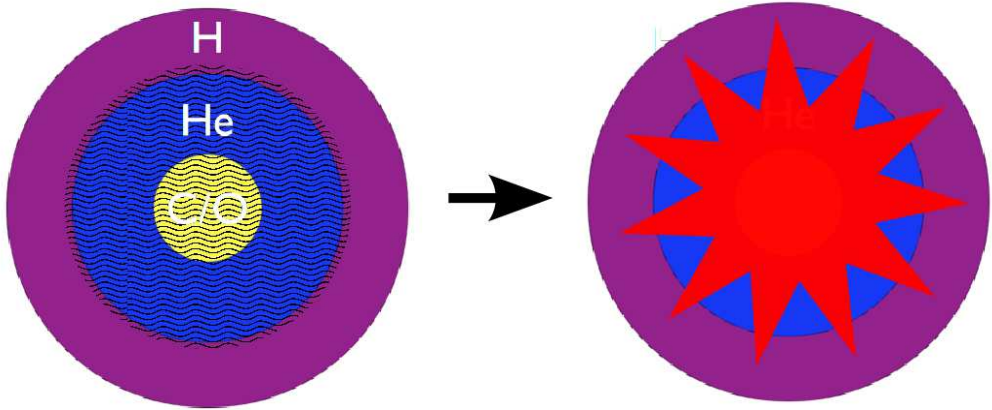}
\caption{ After central helium burning, the helium core of very massive star becomes highly 
convective (wavy mesh) and radiation energy is converted into electron and positron pairs at its 
oxygen core;  This results in an implosion that ignites the oxygen and silicon 
explosively. The energy released from burning eventually blows up the star. 
\label{psn_cartoon}}
\end{center}
\end{figure}

Here our thermonuclear supernovae refer to  those from very massive stars. They are totally 
different from the Type-Ia explosions.  Very massive stars with initial masses of $150-260\,\Msun$ 
develop oxygen cores of $\gtrsim$ $50\,\Msun{}$ after central carbon burning 
\citep{barkat1967,heger2002}.  At this point, the core reaches sufficiently high
temperatures ($\sim 10^9\,\K$) and at relatively low densities ($\sim10^6\,\gcc$) to
favor the creation of electron-positron pairs (high-entropy hot
plasma).  The pressure-supporting photons turn into the rest masses
for pairs and soften the adiabatic index $\gamma$ of the gas
below a critical value of $\frac{4}{3}$, which  causes a dynamical
instability and triggers rapid contraction of the core.  During
contraction, core temperatures and densities swiftly rise, and oxygen
and silicon ignite, burning rapidly.  This reverses the preceding
contraction (enough entropy is generated so the equation of state
leaves the regime of pair instability), and a shock forms at the outer
edge of the core.  This thermonuclear explosion, known as a
pair-instability supernova (PSN), completely disrupts the star with
explosion energies of up to $10^{53}\,\erg$, leaving no compact
remnant and producing up to $50\,\Msun\,\Ni$. Figure~\ref{psn_cartoon}
illustrates  the stellar structure of pre-psn and its explosion.

\begin{figure}[h]
\begin{center} 
\subfigure[Evolution of radial velocity profiles]{\label{vel_prof}\includegraphics[width=0.5\textwidth]{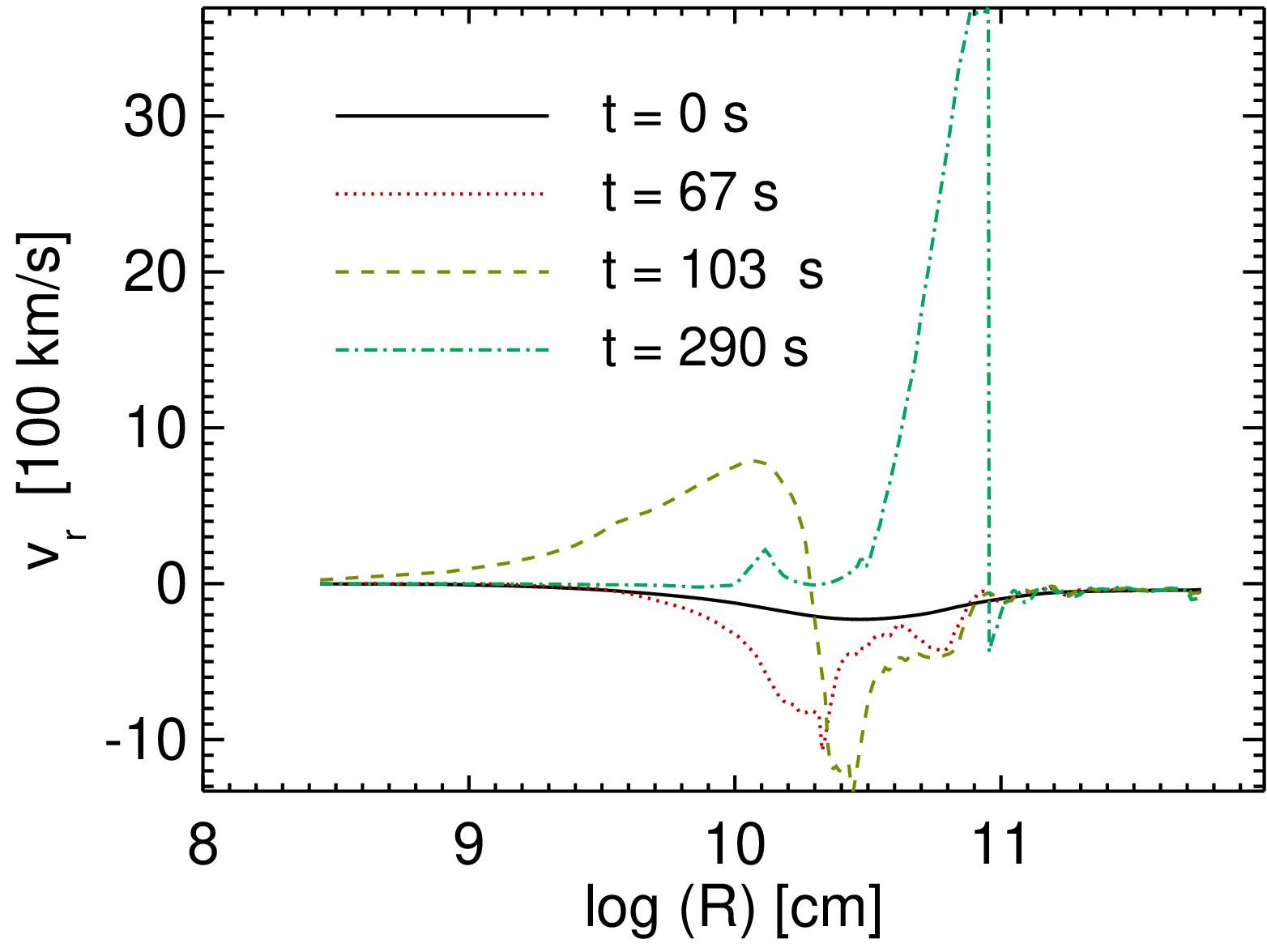}}                
\subfigure[A PSN explosion]{\label{psn} \includegraphics[width=0.4\textwidth]{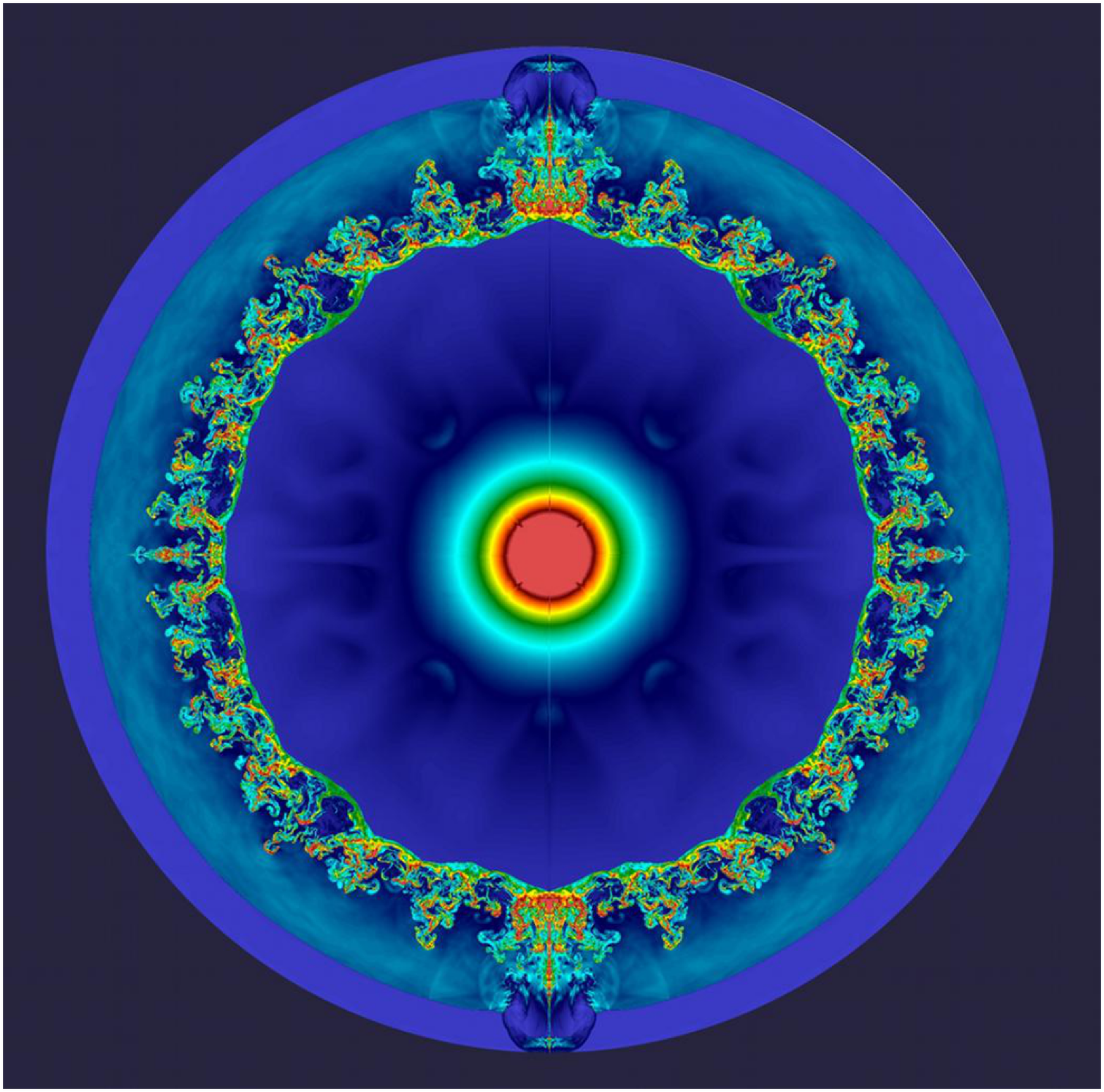}}   
\caption{(a) The in-falling velocities of collapsing core begin at about $200\,\km\sec^{-1}$. 
After the explosion occurs, a strong shock is launched and will propagate until 
it breaks out from the stellar surface.   (b) The fluid instabilities generated during the explosion evolve into a 
large spatial scale and generated a significant mixing. }
\end{center}
\end{figure}

\CASTRO{} \citep{ann2010,zhang2011} is a massively parallel, multidimensional Eulerian adaptive 
mesh refinement (AMR) radiation-hydrodynamics code for astrophysical applications.  Its time 
integration of the hydrodynamics equations is based on a higher-order, unsplit Godunov scheme.  
Block-structured AMR with subcycling in time applies high spatial resolution to where it is needed
most.  We use the Helmholtz equation of state \citep{timmes2000} with density, temperature, and 
elemental abundances;  it includes contributions by non-degenerate and degenerate relativistic 
and non-relativistic electrons, electron-positron pair production, ions and radiation.  The gravitational 
field is calculated with a monopole approximation derived from a radial average of the density on the 
multidimensional grid.  We have implemented several reaction networks (7, 13, 19 isotopes) \citep{
kepler,timmes1999} in \CASTRO{} for calculating nucleosynthesis and energetics in thermonuclear 
SNe.  The most comprehensive network includes alpha-chain reactions, heavy-ion reactions, 
hydrogen burning cycles, photo-disintegration of heavy elements, and energy loss by neutrinos.  

\section{Conservative Mapping}
\lSect{mapping}

Since the star is very nearly in hydrostatic equilibrium and we want to conserve total energy, care 
must be taken when mapping its profile from the uniform Lagrangian grid in mass coordinate to the 
new Eulerian spatial grid.   \citet{zingale2002, mocak2009} have also studied mapping 1D initial conditions onto multidimensional
grids.  Different from our scheme, they focus on maintaining the hydrostatic equilibrium setup, because 
hydrostatic equilibrium is required for their simulations such as modelling X-ray bursts on the surface 
of neutron stars. If their initial conditions do not maintain the hydrostatic equilibrium, the strong gravity
of neutron stars can rapidly pull down the burning layers and cause artificial heating which leads to 
problematic results. To construct a hydrostatic equilibrium profile, their mapping can not conserve physical quantities
such as mass or internal energy. Instead, our problems start with initial conditions that are not 
at the hydrostatic equilibrium and the burning time scale is significantly less than the dynamic
time scale of the star. The proper temperature and density profiles are more important for our problems. 
Our method preserves the conservation of quantities such 
as mass and energy on the new mesh that are analytically conserved in the evolution equations.  
Figure~\ref{vel_prof} shows the radial velocity evolution of an example of PSN simulations.
When $t=0$, $V_{\rm r}$ is nonzero which indicates the initial condition is not hydrostatic equilibrium. Figure~\ref{psn} 
shows a PSN explosion right before the shock breaks out of the stellar surface.

Although this reconstruction does not guarantee that the star will be hydrostatic, it is a physically motivated 
constraint and sufficient for our simulations. The algorithm we describe is specific to our models 
but can be easily generalized to mappings of other 1D data to higher dimensional grids.

\subsection{Method}

First, we construct a continuous (C$^0$) function that conserves the physical quantity upon 
mapping onto the new grid.  An ideal choice for interpolation is the volume coordinate $V$, the 
volume enclosed by a given radius from the center of the star.  Then, integrating a density $\rho
_X$ (which can represent mass or internal energy density) with respect to the volume coordinate 
yields a conserved quantity $X$
\begin{equation}
X=\int _{V_1}^{V_2}\rho_X\,\mathrm{d}V,
\end{equation}
such as the total mass or total internal energy lying in the shell between $V_1$ and $V_2$.

\begin{figure}[h]
\begin{center} 
\includegraphics[scale=0.5]{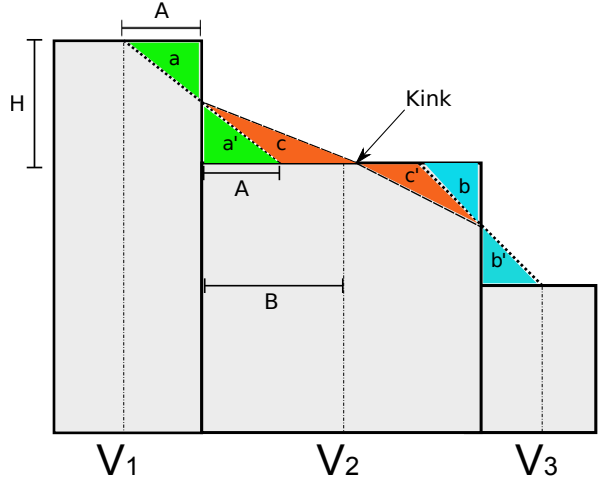}
\caption{Constructing a piecewise-linear conservative profile:  The rectangular bins represent the
original 1D profile.  The areas of different colors represent conserved quantities such as mass and 
internal energy.  The conservative profile connects adjacent bins and makes a =a$'$ = 
$\frac{1}{4} H \times$ min(A,B), c = c$'$, and b = b$'$.  Note that uniform zones in mass (Lagrangian coordinate)  
lead to nonuniform bins in volume coordinate, as shown above.\label{mappingp}}
\end{center}
\end{figure}

Next, we define a piecewise linear function in volume $V$ that represents the conserved quantity 
$\rho_X$, preserves its monotonicity (no new artificial extrema), and is bounded by the extrema 
of the original data.  The segments are constructed in two stages.  First, we extend a line across 
the interface between adjacent zones that either ends or begins at the center of the smaller of the 
two zones, as shown in Figure \ref{mappingp} (note that uniform zones in mass coordinate do not 
result in uniform zones in $V$).  The slope of the segment is chosen such that the area trimmed 
from one zone by the segment (a and b) is equal to the area added under the segment in 
the neighboring bin (a$'$=a and b$'$=b).

If the two segments bounding a and a$'$, and b and b$'$ are joined together by a third in the 
center zone in Figure \ref{mappingp}, two ``kinks'', or changes in slope, can arise in the interpolated 
quantity there; plus, the slope of the flat central segment is usually a poor approximation of the 
average gradient in that interval.  We therefore construct two new segments that span the entire 
central zone and connect with the two original segments where they cross its interfaces, as shown 
in Figure \ref{mappingp}.  The new segments join each other at the position in the central bin where 
the areas c and c$'$ enclosed by the two segments are equal (note that they typically have 
different slopes).  After repeating this procedure everywhere on the grid, each bin will be spanned 
by two linear segments that represent the interpolated quantity $\rho_X$ at any $V$ within the bin 
and have no more than one kink in $\rho_X$ across the zone.  Our scheme introduces some 
smearing (or smoothing) of the data, but it is limited to at most the width of one zone on the original 
grid. Other approaches might be the use of a parabolic reconstruction, such as that described by the PPM \citep{colella1984}, ENO \citep{harten1987}, and WENO \citep{liu1994} schemes. However, these schemes aim mainly for problems with piecewise smooth solutions containing discontinuities. Most models of 1D stellar evolution before their supernova explosions do not contain the discontinuities in the profiles of physical quantities such as density and temperature. Hence our scheme offers a simpler and more effective implementation for the conservative profile reconstruction.

The result of our interpolation scheme is a piecewise linear reconstruction in $V$ of the original 
profile in mass coordinate for which the quantity $\rho_X$ can be determined at any $V$, not 
just the radii associated with the zone boundaries in the Lagrangian grid.  We show this profile as a 
function of the radius associated with the volume coordinate $V$ for a zero-metallicity 200 \Msun{} star with 
$r$ $\sim$ $2 \times 10^{13}$ cm from \KEPLER{} \citep{heger2002,heger2010}.

\begin{figure}[h]
\begin{center} 
\includegraphics[scale=0.7]{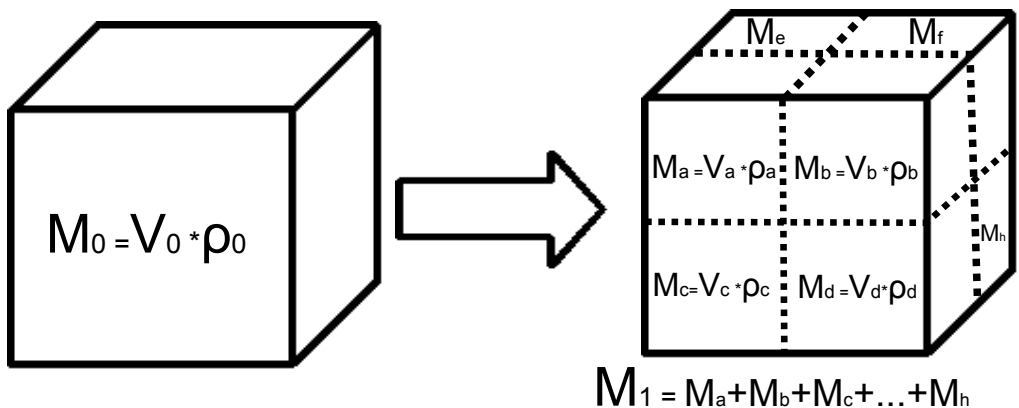} 
\caption{Volume subsampling:   We first use the center of volume element (V$_0$) to obtain its 
density $\rho_0$ from the conservative profile and then calculate its mass M$_0$ = V$_0\times 
\rho_0$.  We then partition the original volume element as shown and calculate the mass of each 
subvolume in the same manner as M$_0$.  We obtain M$_1$ by summing over eight subvolumes 
M$_a$,M$_b$,M$_c$,...,M$_h$.  We then compare M$_1$ and M$_0$; if their relative error is greater than
some predetermined tolerance the process is recursively repeated until $|$(M$_i$-M$_{i-1}$)/M$_{i}$$|$ 
is less than $10^{-4}$.  \label{vol}}
\end{center}
\end{figure}

We populate the new multidimensional grid with conserved quantities from the reconstructed stellar 
profiles as follows.  First, the distance of the selected mesh point from the center of the new grid is 
calculated.  We then use this radius to obtain its $V$ to reference the corresponding density in the 
piecewise linear profile of the star.  The density assigned to the zone is then determined from 
adaptive iterative subsampling.  This is done by first computing the total mass of the zone by 
multiplying its volume by the interpolated density.  We then divide the zone into equal subvolumes 
whose sides are half the length of the original zone.  New $V$ are computed for the radii to the 
center of each of these subvolumes and their densities are again read in from the reconstructed 
profile.  The mass of each subvolume is then calculated by multiplying its interpolated density by its 
volume element (see Figure \ref{vol}).  These masses are then summed and compared to the mass 
previously calculated for the entire cell.  If the relative error between the two masses is larger than 
the desired tolerance, each subvolume is again divided as before, masses are computed 
for all the constituents comprising the original zone, and they are then summed and compared to the 
zone mass from the previous iteration.  This process continues recursively until the relative error in 
mass between the two most recent consecutive iterations falls within an acceptable value, typically 
10$^{-4}$.  The density we assign to the zone is just this converged mass divided by the volume of 
the entire cell.  This method is used to map internal energy density and the partial densities of the 
chemical species to every zone on the new grid.  The total density is then obtained from the sum of 
the partial densities; pressure, and temperature in turn are determined from the equation of state.  
This method is easily applied to hierarchy geometry of the target grid.  

\subsection{Results}

We port a 1D stellar model from  \KEPLER{} into \CASTRO{} to verify that our mapping is 
conservative.  As an example, we use a 200 \Msun{} zero-metalicity pre-supernova star.

\begin{figure}[h]
\begin{center} 
\includegraphics[scale=0.55]{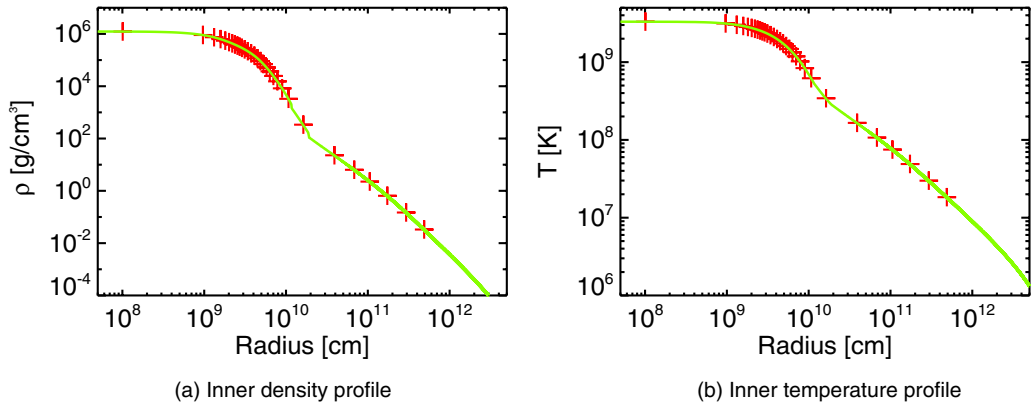}
\caption{Inner density and temperature profiles of a 200  \Msun{} presupernova:  Our piecewise linear profiles 
(green lines) fit their original \KEPLER{} model (red crosses) very well.  Since we map internal energy (a 
conserved quantity) rather than temperature, we calculate $T$ from the equation of state using the density, 
element abundance, and internal energy.  
\label{profile}}
\end{center}
\end{figure}

As we show in Figure \ref{profile}, our piecewise linear fits to the \KEPLER{} data reproducing the 
original stellar profile.  Because our fits smoothly interpolate the block histogram structure of the 
\KEPLER{} bins (especially at larger radii), they reduce the number of unphysical sound waves 
that would have been introduced in \CASTRO{} by the discontinuous interfaces between these 
bins in the original data\footnote[1]{1D data usually provides zone-averaged values, 
hence a continuous and conservative profile needs to be reconstructed from zone-averaged values.}.  
The density profile is key to the hydrodynamic and gravitational evolution 
of the explosion, and the temperature profile is crucial to the nuclear burning that powers the 
explosion.

\begin{figure}[h]
\begin{center} 
\includegraphics[scale=0.5]{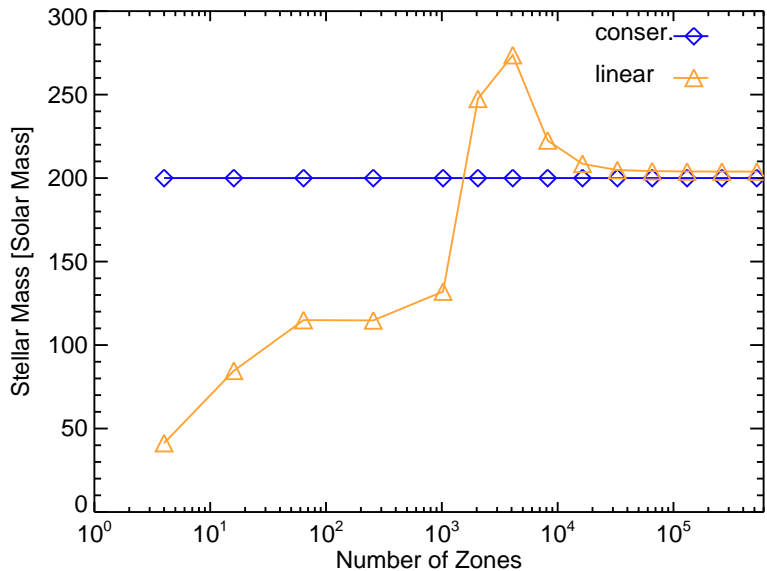}
\caption{Total mass of the star on the 1D \CASTRO{} grid as a function of number of zones:   Conservative 
mapping (blue) preserves the mass of the star at all resolutions, while linear interpolation (orange) 
converges to 200 \Msun{} at a resolution of $\sim$ a few $\times 10^4$, when the grid begins to 
resolve the core of the star ($\sim 10^9$ cm).  Even at very high resolutions, the results of linear 
interpolation are still off by a few percent from the targeted mass and start to be saturated at 
$\sim 10^5$ zones because the linear interpolation profile is not a conservative one.  \label{map1d}}
\end{center}
\end{figure}

We first map the profile onto a 1D grid in \CASTRO{} and plot the mass of the star as a function 
of grid resolution in Figure \ref{map1d}.  The mass is independent of resolution for our conservative 
mapping because we subsample the quantity in each cell prior to initializing it, as described above.  
In contrast, the total mass from linear interpolation is very sensitive to the number of grid points but 
does eventually converge when the number of zones is sufficient to resolve the core of the star, in 
which most of its mass resides.  

\begin{figure}[h]
\begin{center} 
\subfigure[2D Mapped Mass of Entire Star]{\label{map2d_s}\includegraphics[width=0.45\textwidth]{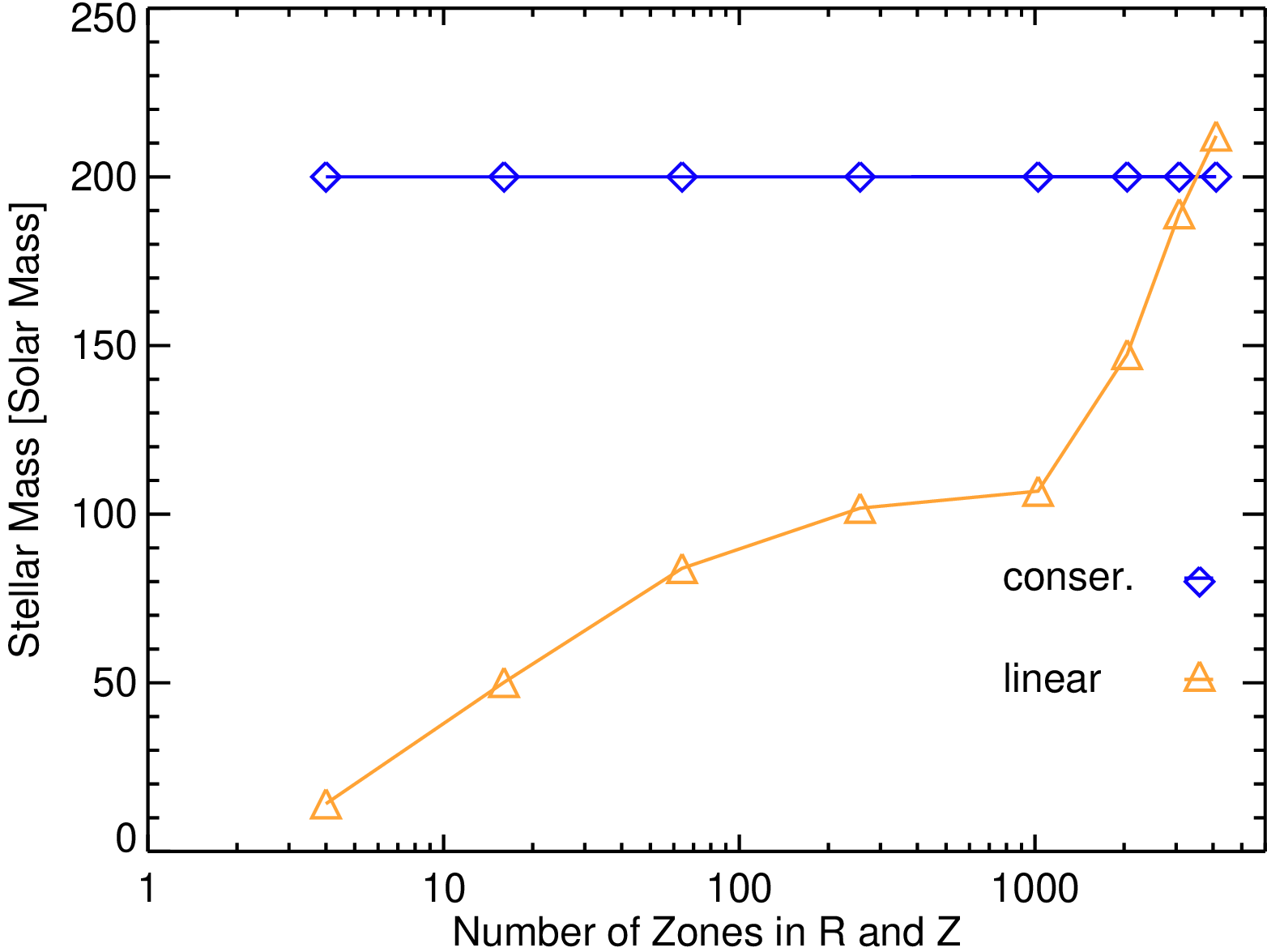}}                
\subfigure[2D Mapped Mass of Helium Core]{\label{map2d_c} \includegraphics[width=0.45\textwidth]{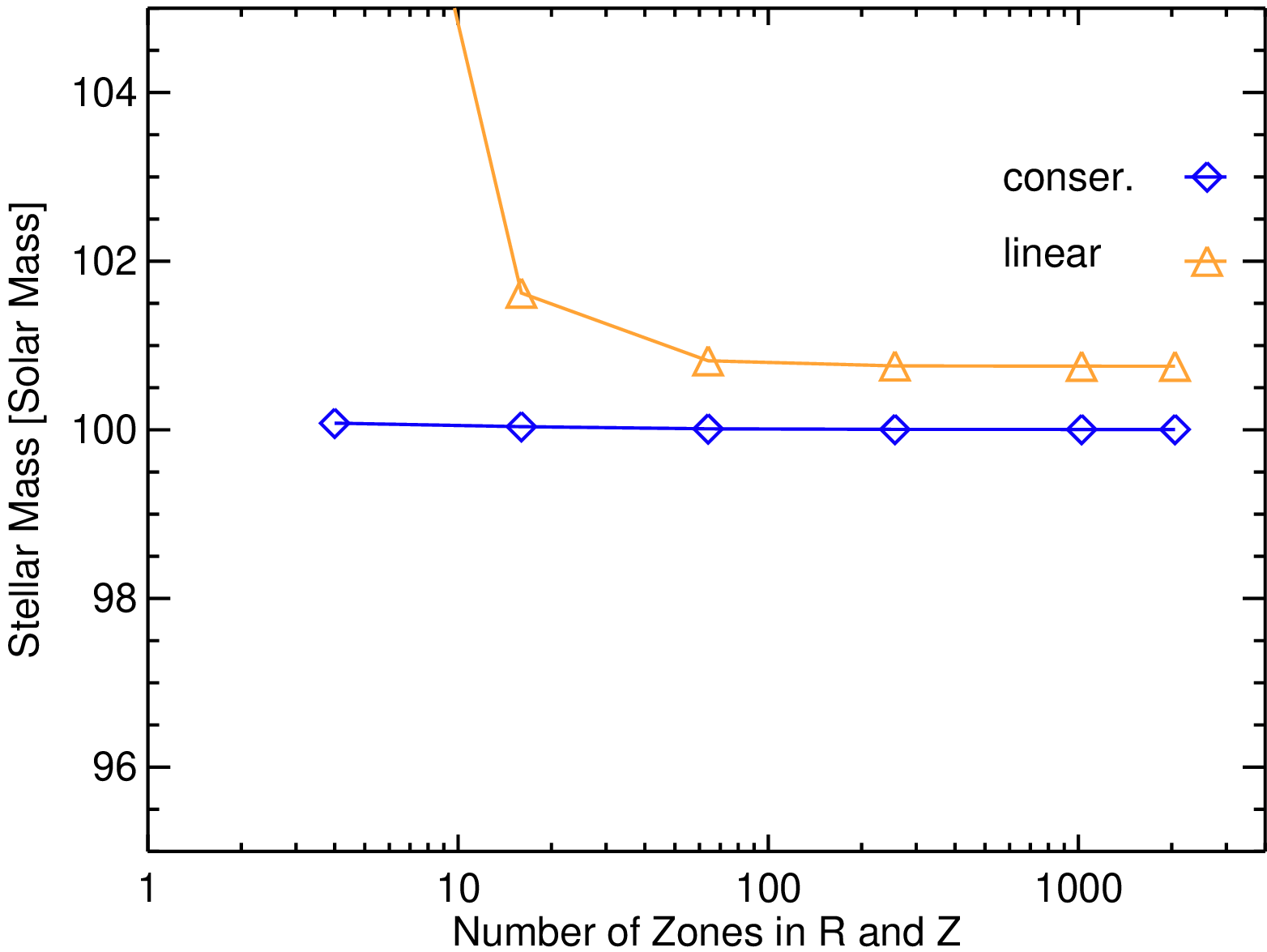}}   
\caption{(a) Total mass of the star on the new 2D \CASTRO{} grid as a function of number of zones in both 
$r$ and $z$:  Conservative mapping (blue) recovers the mass of the star at all resolutions and 
linear interpolation (orange) approaches 200 \Msun{} at a resolution of $\sim$ $2048^2$.  (b) 
Total mass of the He core on the 2D \CASTRO{} grid as a function of  number of zones in both $r$ and $z$:  
Conservative mapping (blue) preserves its original mass at all resolutions while linear interpolation 
(orange) begins to converge to 100 \Msun{} at a resolution of $64^2$, but it is still off by $\sim 1 \%$ 
even as the resolution approaches $\sim 2048^2$ because the linear interpolation profile is, by nature, not conservative.}
\end{center}
\end{figure}

We next map the \KEPLER{} profile onto a 2D cylindrical grid ($r,z$) and a 3D cartesian grid 
($x,y,z$) in \CASTRO{}.  The only difference between mapping to 1D, 2D, and 3D is the form of 
the volume elements used to subsample each cell,  which are $4\pi r^2 dr$, $2\pi r  dr dz$, $dx 
dy dz$, respectively.  We show the mass of the star as a function of resolution in Figure 
\ref{map2d_s}.  Conservative mapping again preserves its mass at all grid resolutions.  In 2D, 
more zones are required for linear interpolation to converge to the mass of the star.  To further 
verify our conservative scheme, we map just the helium core of the star ($\sim$ 100 $\Msun$ 
with $r \sim$ 10$^{10}$ cm) onto the 2D grid.  The helium core is crucial to modeling 
thermonuclear supernovae because it is where explosive burning begins.  We show its mass as 
a function of resolution in Figure \ref{map2d_c}.  We again recover all the mass of the core at all 
resolutions because linear interpolation overestimates the mass by at least $\sim$ 1 $\%$, even with 
large numbers of zones.  

Because of the property of reconstruction, conservative mapping is still valid  in 3D but requires much more 
computational time to subsample each cell to convergence. Furthermore, an impractical number of zones 
is needed for linear interpolation to reproduce the original mass of the star. So we do not show the comparison
of 3D models. We note that our method also works with AMR grids because both $V$ and the interpolated quantities can 
be determined, and subsampling can be performed on every grid in the hierarchy.  For the given domain, the 
results of conservative mapping are independent of the levels of AMR.     

\section{Initial Perturbation}
\lSect{perb}

Seeding the pre-supernova profile of the star with realistic perturbations may be important to 
understanding how fluid instabilities later erupt and mix the star during the explosion.  Massive 
stars usually develop convective zones prior to exploding as SNe \citep{woosley2002,heger2002}.  
Multidimensional stellar evolution models suggest that the fluid inside the convective regions can 
be highly turbulent \citep{porter2000,arnett2011}.  However, in lieu of the 3D stellar evolution 
calculations necessary to produce such perturbations from first principles, multidimensional 
simulations are usually just seeded with random perturbations.  In reality, if the star is convective 
and the fluid in those zones is turbulent \citep{davidson2004}, a better approach is to imprint the 
multidimensional profiles with velocity perturbations with a Kolmogorov energy spectrum \citep{frisch1995}.

Next we describe our scheme for seeding 2D and 3D stellar profiles with turbulent perturbations and 
present stellar evolution simulations with \CASTRO{} with these profiles.  In our setup, the 
perturbations have the following properties:

\begin{enumerate}
\item The perturbations are imprinted in the gas velocity, and their net momentum flux must be 
zero.  Because the initial perturbations only play as seeds for any fluid instabilities and 
we want to minimize the overall impact of perturbed velocities to the dynamics of star. 
$\nabla \cdot (\rho v) =0$ may not be fulfilled locally. So strictly speaking, the perturbed 
velocity field is not solenoidal.

\item They are seeded in convectively unstable regions with a velocity spectrum $V(k) \sim  k^
{-5/6}$, where $k$ is the wave number and the power index $-5/6$ is for a Kolmogorov spectrum with an 
assumption of constant density. We assume a low Mach number convection, which implies that 
the fluid can be approximated as incompressible, which leads to the 
density contrast of convective bubbles being small. The 1D MLT of our 
models also suggest that convective velocities are subsonic.

\end{enumerate}

\subsection{2D Perturbation}
We first consider the mapping onto a polar coordinate grid in $r$ and $\theta$.  To enforce zero 
net momentum and the boundary conditions in the simulation, we define a new variable $x = 1+
\cos\theta$ instead of using $\theta$.  The momentum flux of a density $\rho$ and velocity $v_r$  
is then
\begin{equation}
\int_0^\pi \! 2\pi r^2  \rho v_r \sin\theta \, \mathrm{d} \theta  = \int_0^2 \! 2\pi r^2 \rho v_r  \, \mathrm{d}x = 0
\end{equation}
if $v_r$  has the form $\cos(2\pi n x)$, where $n$ is an integer.  When $\theta$ = 0, $\pi$ (the 
boundaries of a 2D grid), $x$ = 2, 0 yields the maximum values for $v_r$ that satisfy the 
boundary conditions in 2D cylindrical coordinates in \CASTRO{}.  There 
are two physical scales that constrain the wavelength of the perturbation in $r$.  Based on the mixing 
length theory \citep{cox1965}, the eddy size of turbulence is $\alpha \times H_{\rm p}$; $\alpha$ is the mixing 
length parameter, and $H_{\rm p}$ is the pressure scale height.  Here, we set $\alpha$ = 1.0.  Since the perturbation is 
only seeded in the convective zones, it is confined inside domain $D_c = r_{\rm u}$ - $r_{\rm b}$, where $r_{\rm u}$ 
and $r_{\rm b}$ are its upper and lower boundaries.  The maximum wavelength of the perturbation must be smaller 
than $D_{\rm c}$ and $H_{\rm p}$.  Inside a convective zone, we define a new variable, $y = \int_{r_{\rm b}}^{r} \frac{{\rm d}r}{H_{\rm p}(r)}$.  
We also define two oscillatory functions in $x$ and $y$ to generate 
the circular patterns that mimic the vortices of a turbulent fluid.  Since the fluid inside the convective 
zone is turbulent, its energy spectrum is $E(k) \sim  k^{-5/3}$.  Assuming a constant density, the 
corresponding velocity spectrum is $V(k)\sim  k^{-5/6}$.  The perturbed velocity then has the form,
\begin{equation}
\begin{array}{l}
V_{{\rm perb},r}(x,y) = -\sum\limits_{a}\sum\limits_{b} V_{\rm p}\cdot\cos(2\pi ax)\cdot\cos(2\pi by+\alpha_b),   \\
V_{\rm perb,\theta}(x,y) =\sum\limits_{a}\sum\limits_{b} V_{\rm p}\cdot\sin(2\pi ax)\cdot\cos(2\pi by+\alpha_b),\\
V_{\rm p} = V_{\rm d} (r) b^{-5/6},
\end{array}
\label{eq:perb2d-1}
\end{equation}
where $V_{{\rm perb},r}$ and $V_{\rm perb,\theta}$ are the perturbed velocities in the $r$ and $\theta$ 
directions, and $a$ and $b$ are angular and radial wavenumbers.  1D models provide only the 
information of convective velocities, $V_{\rm d}(r)$ along the radial direction, which can be treated as average velocities
of angular directions, so we scale the amplitude of the perturbed velocity based on the radial wavenumber $b$.  Besides, we use the oscillatory functions
for constructing the eddy-like pattern of perturbed velocity field which 
provides an alternative way to angularly distribute our perturbation. 
This is based on a physically motivated way, which is more realistic 
than purely random perturbations.
In a realistic turbulent follow, $V_{\rm p}$ of 
eddies should depend only on the scale of physical
length without preferred direction. In our setup, we simplify
the implementation by constraining the length scale only in the 
radial direction. 
Our oscillatory functions then decompose $V_{\rm p}$ in 
$r$, $\theta$, and $\phi$  directions.
We also use a random phase, $\alpha_b$, to smooth out numerical discontinuities caused by the perturbed 
modes while summing.  Equations (\ref{eq:perb2d-1}) by construction satisfy $ V_{\rm perb,\theta}(r,\theta) = 0$ 
when $\theta = 0$ and $\pi$, the boundary conditions in $\theta$ on the 2D grid.  
The assumption of no overshooting makes $V_{{\rm perb},r} = 0$ at the boundaries of convective zones, so we set 
$V_{{\rm perb},r} = 0$ at boundaries. The ultimate wavenumbers  of $a$ and $b$ are 
also limited by $D_c$, $H_p$, and the resolution of simulation, $H_{\rm res}$.  

\subsection{3D Perturbation}
In 3D, we use spherical coordinates, $r$, $\theta$, and $\phi$.  Similar to 2D, we construct an oscillatory 
function for ($\theta,\phi$) by using spherical harmonics, $Y_{l,m}(\theta,\phi)$, where $l$ and 
$m$ are the wavenumbers in $\theta$ and $\phi$.  If the velocities are in the form of $Y_{l,m}(\theta,\phi)$, they 
automatically conserve momentum flux while summing all the modes $l,m$.  In the radial direction, we use $\cos(cy)$, where $c$ is the wavenumber in the radial direction and $y$ is as defined in 2D.  The perturbation then has the following form: 

\begin{equation}
\begin{array}{l}
V_{{\rm perb},x}(r,\theta,\phi) = V_{\rm perb}\sin(\theta)\cos(\phi),\\
V_{{\rm perb},y}(r,\theta,\phi) = V_{\rm perb}\sin(\theta)\sin(\phi),\\
V_{{\rm perb},z}(r,\theta,\phi) = V_{\rm perb}\cos(\theta), \\
V_{\rm perb} = \sum\limits_{c}\sum\limits_{l}\sum\limits_{m}V_{\rm p} \cdot Y_{l,m}(\theta+\omega_{lm},\phi+\omega_{lm})\cdot\cos(2\pi cy+\lambda_c),  \\
 V_{\rm p} = V_{\rm d}(r)  c^{-5/6},
\end{array} 
\label{eq:perb3d}
\end{equation}
where $V_{{\rm perb},x}$,  $V_{{\rm perb},y}$,  $V_{{\rm perb},z}$ are the perturbed velocities in the $x$, $y$, and $z$ 
directions.  We sum over the modes, applying random phases $\omega_{lm}$ and $\lambda_c$ to 
smooth out numerical discontinuities caused by different perturbed modes. Similar to 2D, $V_{\rm p}$ is 
only scaled by radial wavenumber $c$.  Because there are no reflective boundary conditions for 3D, we
only take care of the boundary conditions in radial direction. We again assume there is no overshooting
outside the boundaries of convective zones, so we enforce $V_{\rm perb}$ to zero at boundaries.  

\subsection{Results}
We first initialize perturbations on a 2D grid with a profile that is derived from a 1D \KEPLER{} 
stellar evolution calculation.  The perturbations are confined to regions that are convectively 
unstable \citep{heger2000}.  The magnitude $V_{\rm d}(r)$ of the perturbed velocity adopts the 
diffusion velocity, which is usually $\sim$ $1-10\%$ of the local sound speed.  We again consider 
a zero-metalicity $200\,\Msun$ star in the pre-supernova phase.  This star develops a large 
convection zone that extends out to the hydrogen envelope.  We show the magnitude of the 
perturbed velocity generated by the two oscillatory functions discussed above on our 2D grid in 
Figure \ref{perb_2d}.  The velocity field satisfies the reflecting boundary conditions on the 2D grid 
at $\theta$ = 0 and $\pi$.  In the right panel we show velocity vectors in the selected subregion on 
the left (blue rectangle).  A clear vortex pattern that mimics a turbulent fluid is clearly visible.  Next 
we calculate energy spectrum of perturbed velocity field. We first randomly pick a radial direction
( constant $\theta$ in 2D) or (constant $\theta$ and $\phi$)in 3D) inside the convective zone, perform 
Fourier transfer of $V_{\rm p}$ 
along the radial direction, then calculate its power spctrum. We repeat the same process 
ten times, our final spectrum is obtained by averaging all spectra previously calcuated.
$k$ = $H_{\rm p}/l$, where $H_{\rm p}$ is the pressure scale height and $l$ is the 
physical scale in $r$ direction. Figure~\ref{perb_2dp} shows the energy spectrum of the fluid, which is 
basically a Kolmogorov spectrum $E(k) \sim k^{-5/3}$ except for fluctuations in part caused by the random phases in the sum over 
modes in $r$, and $V_{\rm d}(r)$ is not a constant across the convective region that produces an offset 
in the smaller $k$ region.  The energies would converge to the Kolmogorov spectrum in the limit of large 
$k$, but the maximum $k$ of our simulation is limited to the resolution of the grid.  

We next port our 1D \KEPLER{} model to a 3D grid.  In Figure \ref{perb_3d}, we show a slice of 
the magnitude of the perturbed velocity, which again exhibits the clear cell pattern reminiscent of 
the vortices of a turbulent fluid.  The velocity pattern in 3D is more irregular than in 2D.  We show 
the energy spectrum of the velocity field in Figure \ref{perb_3dp}, which is similar to that of our 2D 
spectrum but with larger fluctuations that are again due to the random phases we assign to each 
spherical harmonic, and the $V_{\rm d}(r)$ is not a constant across the convective region that produces 
an offset in the smaller $k$ region.   
We also check the values of  perturbed velocities whether they are consistent to the $V_{\rm d}(r)$ or not. 
We calculate the variance of radial velocities; $\delta V_r = \langle V(r) -\langle V(r) \rangle \rangle$. 
Figure~\ref{diffv} shows the comparison between {$\delta V_r$ and $V_{\rm d}$ as 
a function of radius. The values of $\delta V_r$ are consistent to the original $V_{\rm d}$. 
 The oscillatory pattern of $\delta V_r$ comes from our formalism Equations~(\ref{eq:perb3d}). 
Above examples demonstrate that our scheme effectively generates turbulent 
fluid perturbations analog to those found in the convective regions of massive 
stars, with the desired velocity patterns and energy power spectra.

We do not claim the models here can fully reproduce the true turbulence found in simulations or laboratories. Unlike previous multidimensional simulations of this kind,
whose initial perturbations were seeded by numerical noises or random perturbations. The scheme here is the first attempt to model the initial perturbations based on a more realistic
setup, where the convective zones of a star play an ideal role for generating perturbations.
These kinetic energy of these perturbations is very small compared with the internal energy of the gas, thus 
it does not interfere with the overall dynamics of the simulations or trigger an artificial ignition.
We seed initial perturbations to trigger the fluid instabilities on multidimensional simulations so 
we can study how they evolve with their surroundings as shown in  Figure~\ref{psn}.  When the fluid instabilities start to evolve 
nonlinearly, the initial imprint of perturbation would be smeared out.  The random perturbations and turbulent
perturbations then give consistent results.  Depending on the nature of problems, the random perturbations might take 
a longer time to evolve the fluid instabilities into turbulence because more relaxation time is required.

\begin{figure}[h]
\begin{center} 
\subfigure[2D Perturbed Velocity]{\label{perb_2d}\includegraphics[width=0.4\textwidth]{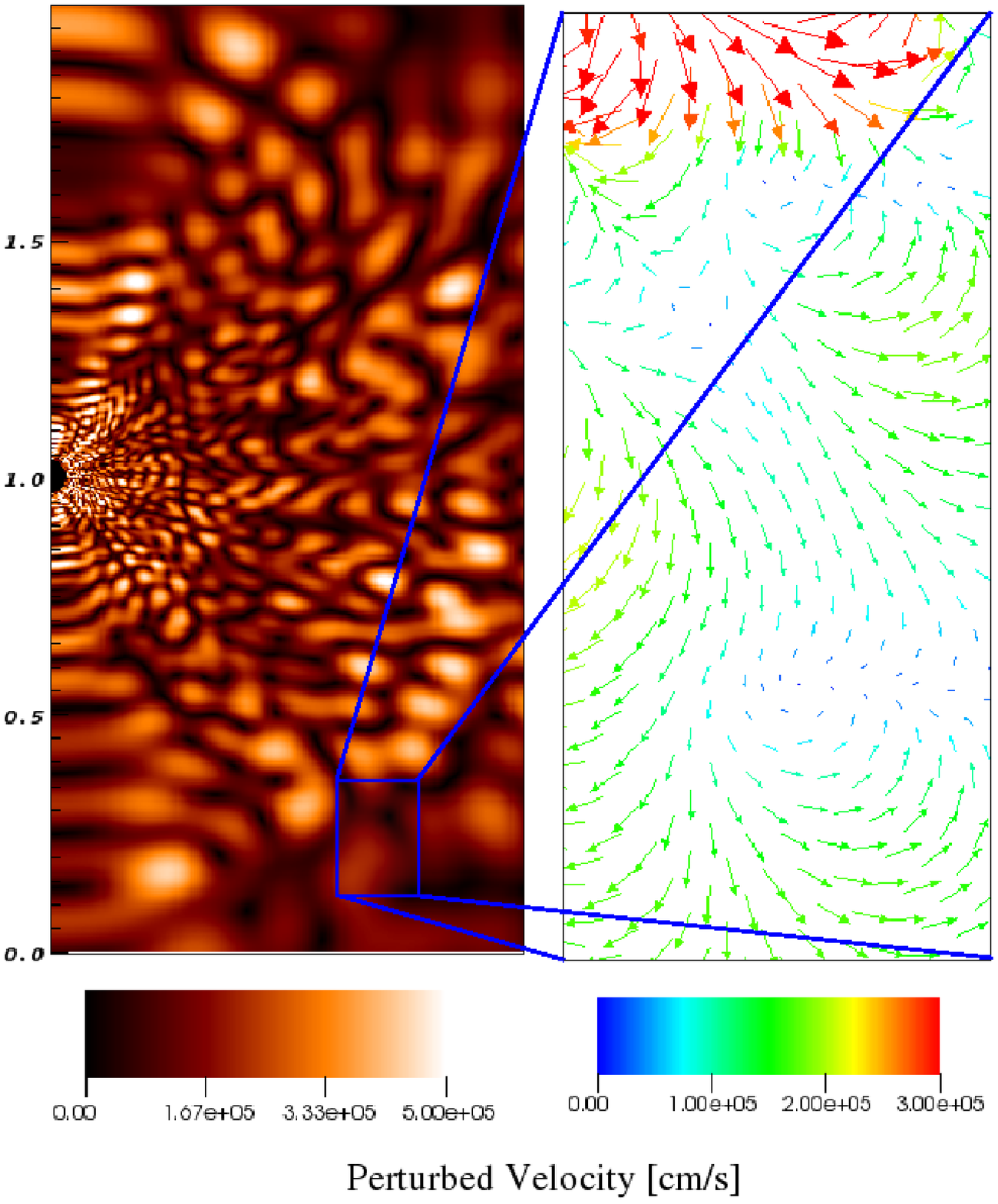}}                
\subfigure[2D Energy Power Spectrum]{\label{perb_2dp} \includegraphics[
width=0.55\textwidth]{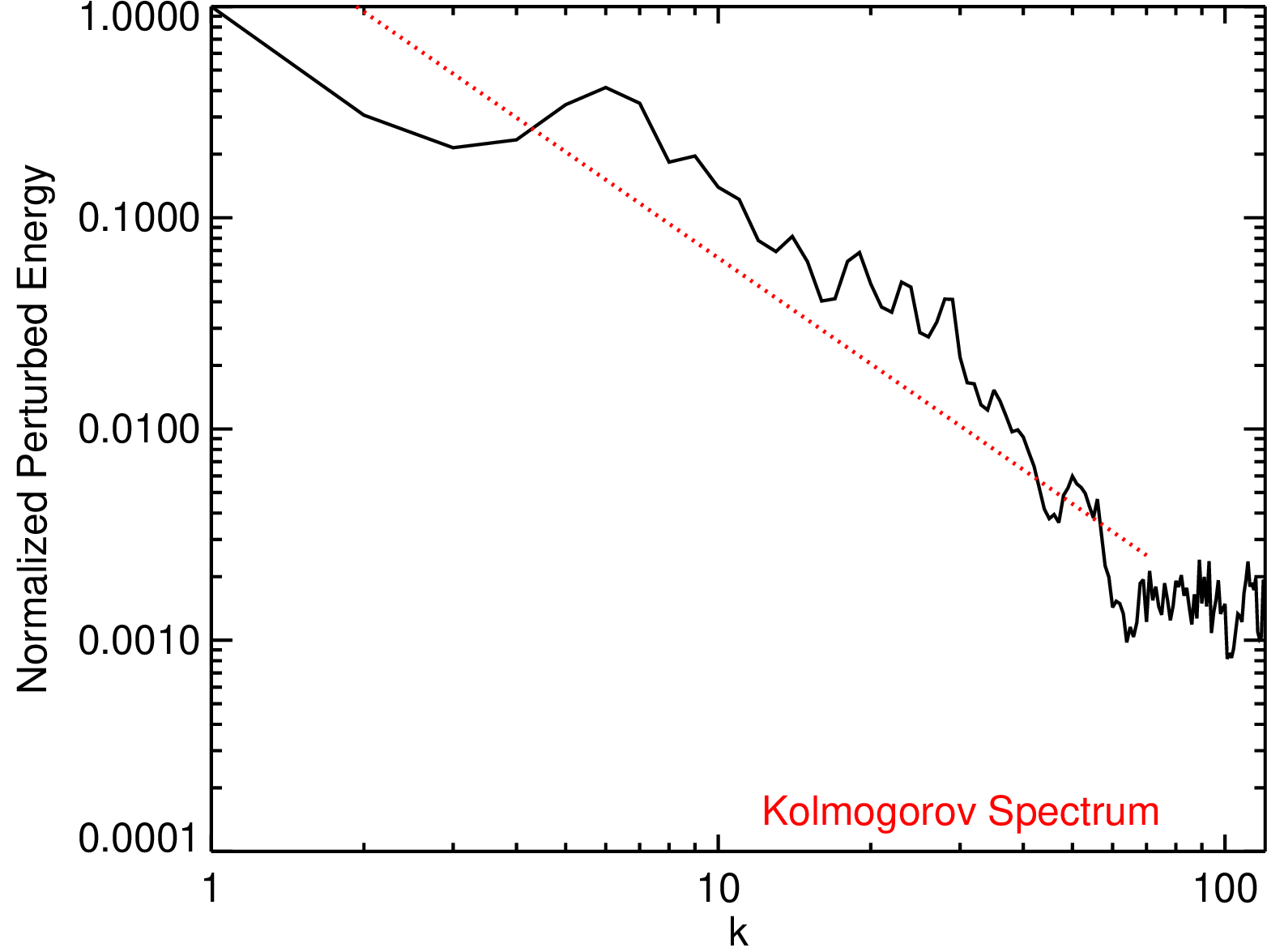}}   
\caption{ (a) 2D perturbed velocity in the interior of the star on physical 
scales $\sim 10^{12}$ cm:  
The closeup is the velocity vector field corresponding to the blue rectangle and exhibits a vortex 
pattern similar to that of a turbulent fluid.  (b) Normalized kinetic energy power spectrum of a 2D 
perturbed field:  The dotted red line is the Kolmogorov spectrum, $ E(k) \sim k^{-5/3}$.  The peak of 
the Kolmogorov spectrum is adjusted to fit the data.  The scale of $H_p$ is equaled to $k = 1$.
The suppressed power at lower $k$ is because of the inhomogeneous 
of $V_{\rm p}(r)$ at larger scales. The decay trend follows $k^{-5/3}$, and the fluctuations are caused by 
the radial oscillatory function with random phases. }
\end{center}
\end{figure}

\begin{figure}[h]
\begin{center} 
\subfigure[3D Perturbed Velocity]{\label{perb_3d}\includegraphics[width=0.35\textwidth]{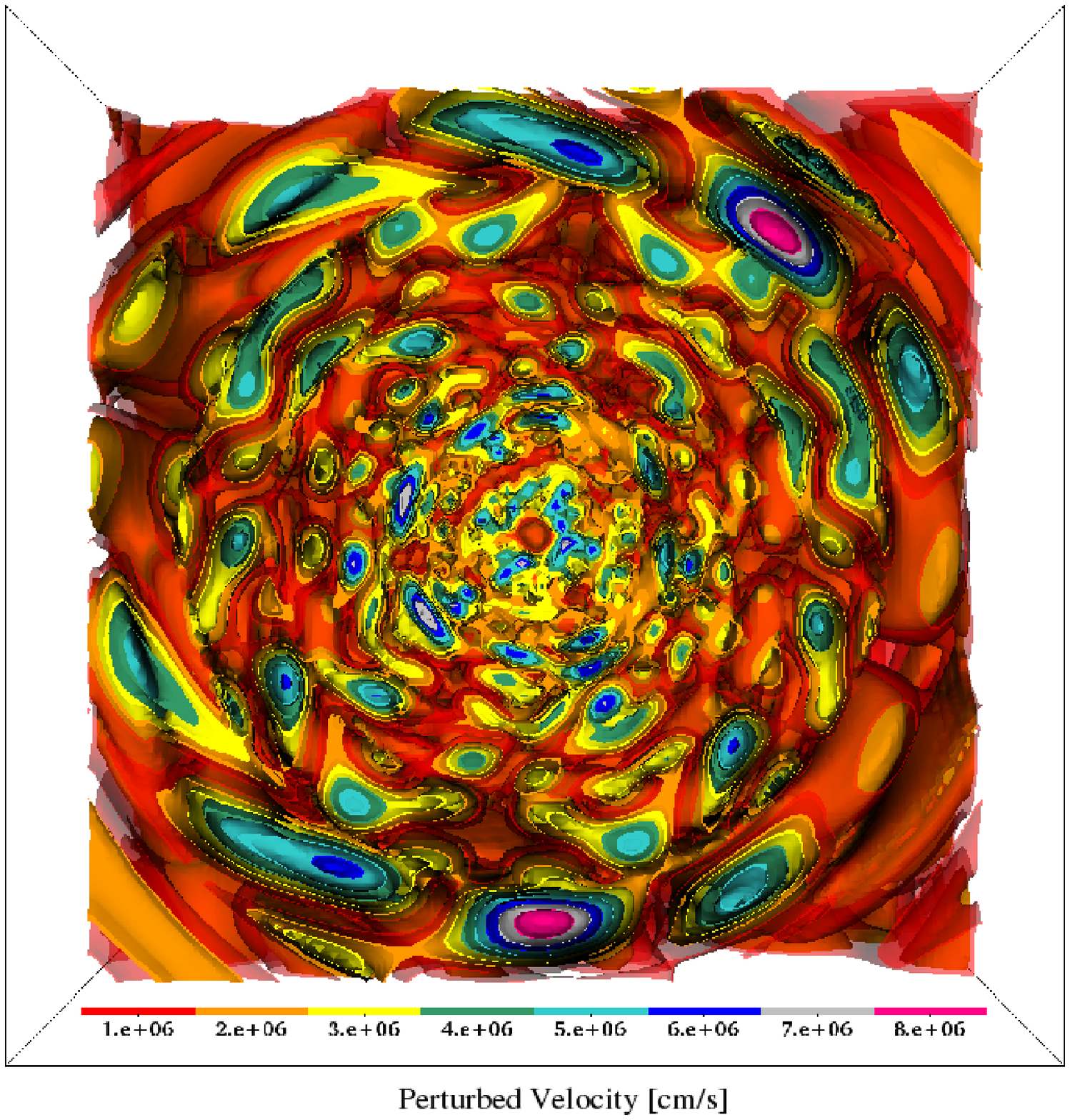}}                
\subfigure[Kinetic Energy Power Spectrum of 3D perturbed field]{\label{perb_3dp} \includegraphics[
width=0.4\textwidth]{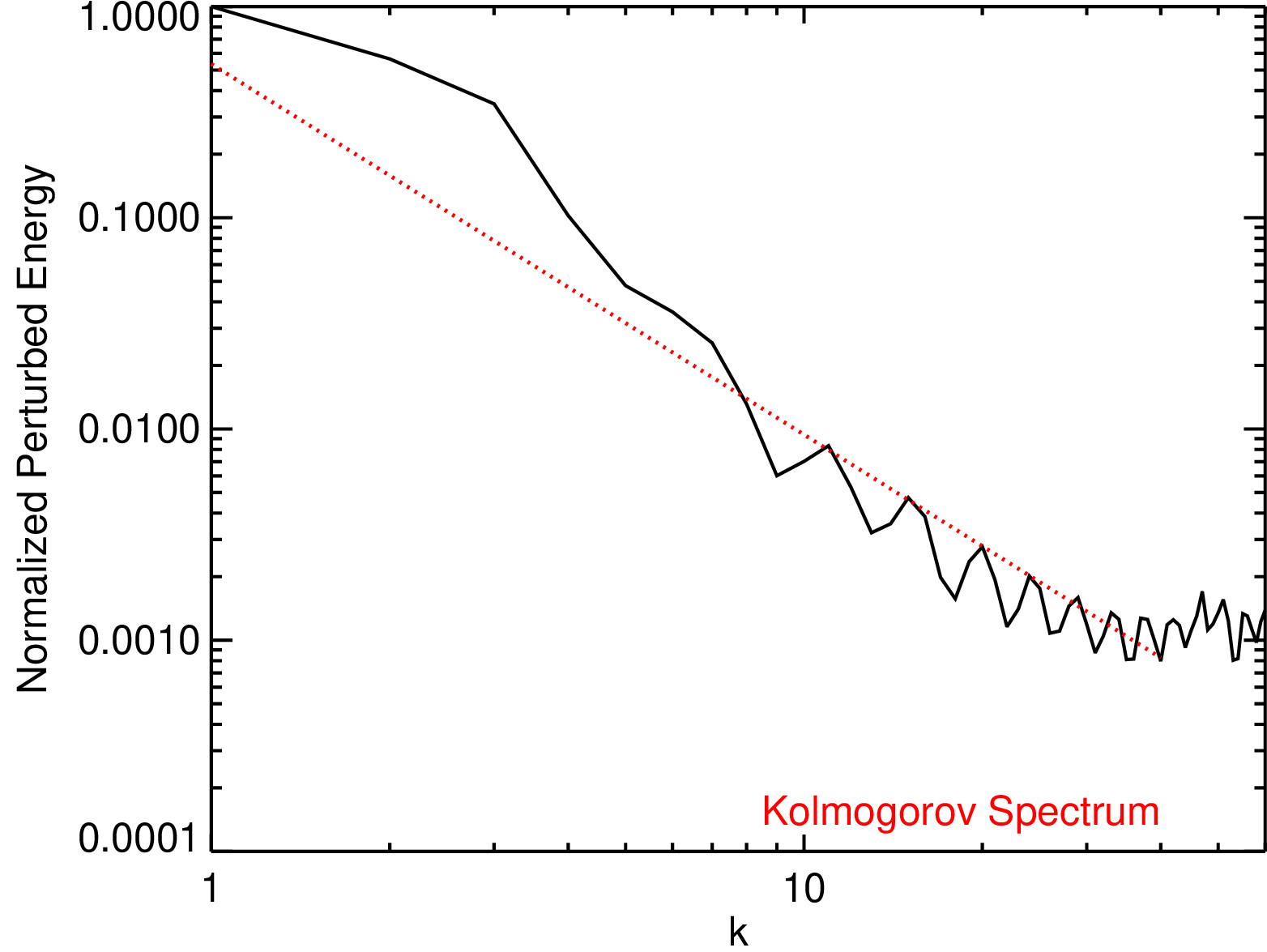}}   
\caption{ (a) 3D perturbed velocity field:  The iso-surfaces show the magnitude of the perturbed 
velocity on physical scales of $10^{12}$ cm.  
(b) Corresponding energy power spectrum:  The dotted red line shows the Kolmogorov spectrum, 
$ E(k) \sim k^{-5/3}$.  The peak of the Kolmogorov spectrum is adjusted to fit the data.  The scale of 
$H_p$ is equaled to $k=1$.  Similar to 2D, the decay trend follows $k^{-5/3}$, and the fluctuations 
are caused by the radial oscillatory function with random phases.}
\end{center}
\end{figure}

\begin{figure}[h]
\begin{center} 
\includegraphics[scale=0.55]{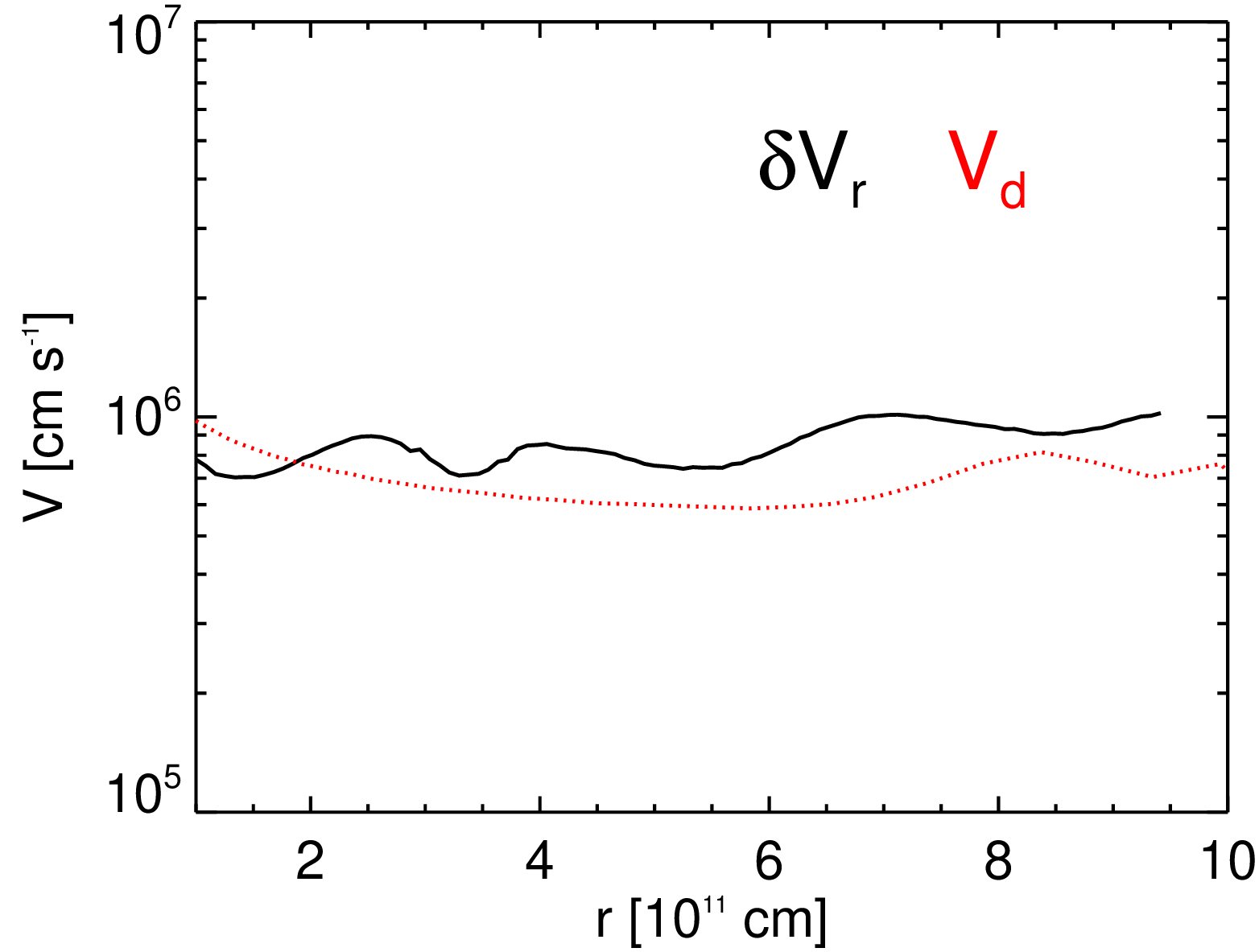}
\caption{$\delta V_r$ and $V_{\rm d}$ as a function $r$ inside the convective zone of the star: The values of $\delta V_r$  inside the convective zone are about $10^{6}\,\cm\,\sec^{-1}$, those are consistent to the values of $V_{\rm d}$, predicted from the 1D MLT theory. The oscillatory 
trend of $\delta V_r$ reflects the original function form of the perturbations.
\label{diffv}}
\end{center}
\end{figure}

\section{Resolving the Early Stages of the Explosion}
 \lSect{resolution}
 In addition to implementing realistic initial conditions and relevant physics for \CASTRO, care must be taken to 
 determine the resolution of multidimensional simulations required to resolve the most important physical scales and yield 
 consistent results, given the computational resources that are available.  We provide a systematic 
 approach for finding this resolution for multidimensional stellar explosions.  
 
 Simulations that include nuclear burning, which governs nucleosynthesis and the energetics of the 
 explosion, are very different from purely hydrodynamical models because of the more stringent
 resolution required to resolve the scales of nuclear burning and the onset of fluid instabilities in 
 the simulations.  Because energy generation rates due to burning are very sensitive to temperature, 
 errors in these rates as well as in nucleosynthesis can arise in zones that are not fully resolved.  
 We determine the optimal resolution with a grid of 1D models in \CASTRO{}.  Beginning with a
 crude resolution, we evolve the pre-supernova star and its explosion until all burning is complete 
 and then calculate the total energy of the supernova, which is the sum of the gravitational energy, 
 internal energy, and kinetic energy.  We then repeat the calculation with the same setup but with a finer 
 resolution and again calculate the total energy of the explosion.  We repeat this process until the 
 total energy is converged.  As shown in Figure \ref{res}, our example of a $200\,\Msun$  presupernova 
 converges when the resolution of the grid approaches $10^8\,\cm$.

\begin{figure}[h]
\includegraphics[scale=0.35]{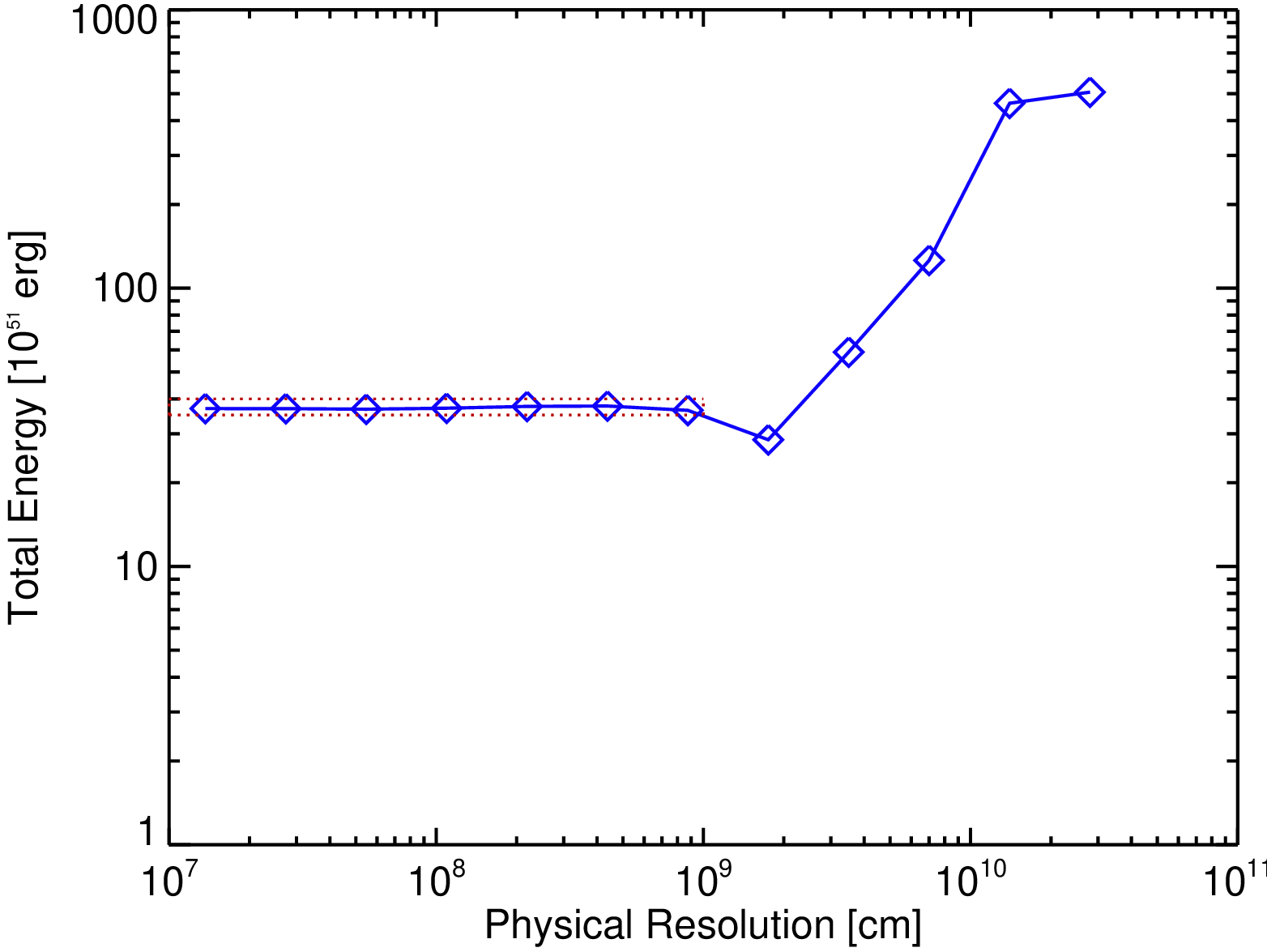}
\includegraphics[scale=0.35]{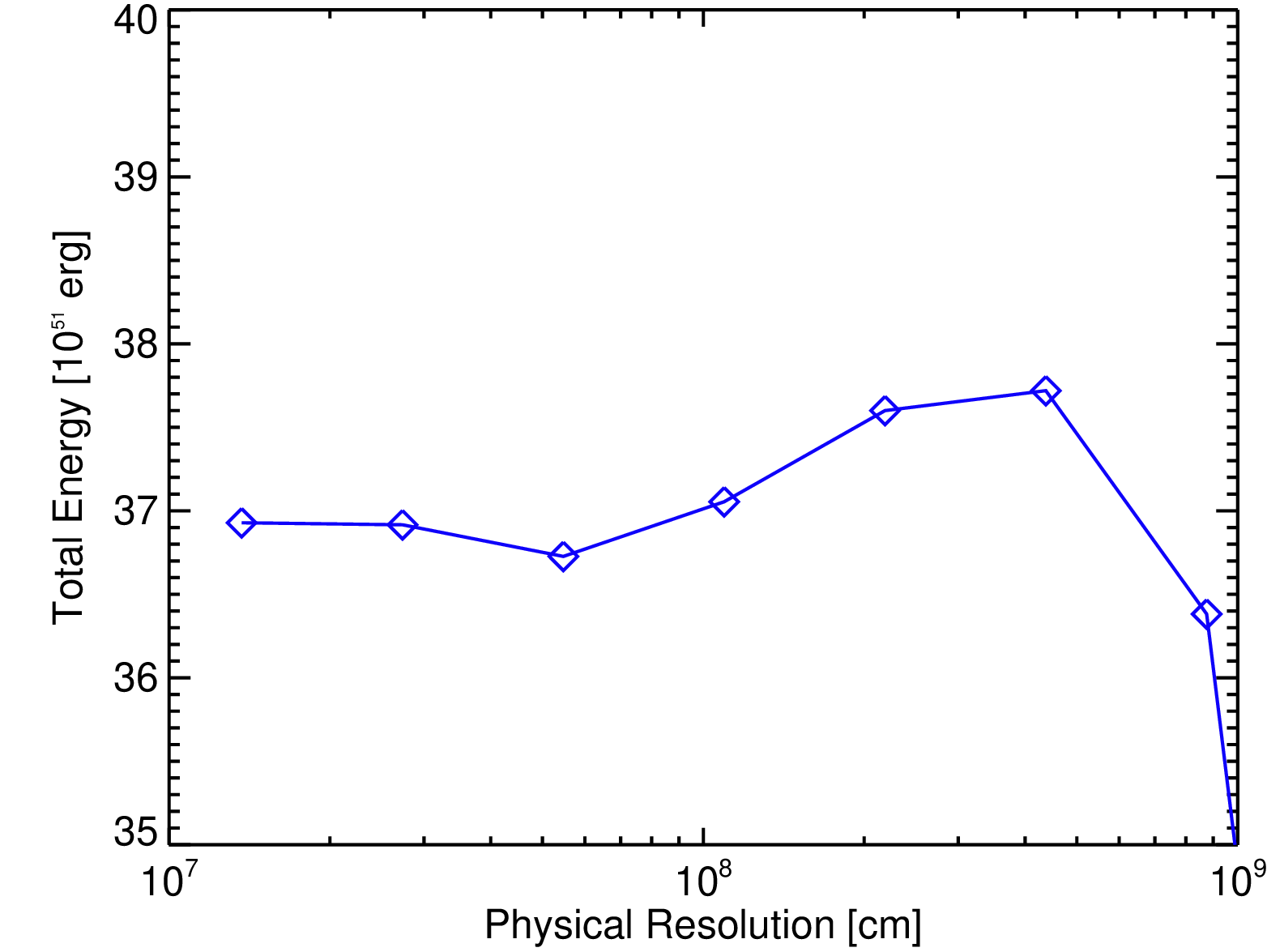}
\caption{Total explosion energy as a function of resolution:  the x-axis is the grid resolution and 
the y-axis is the total energy, defined to be the sum of the gravitational energy, the internal energy, 
and the kinetic energy.  The total energy is converged when the resolved scale is close to $ \Ep{8}$ cm.  
The right panel shows the zoom-in of the red box in left panel.   \label{res}}
\end{figure}

The time scales of burning (${\rm d}t_{\rm b}$) and hydrodynamics (${\rm d}t_{\rm h}$) can be very disparate, 
so we adopt time steps of $ min({\rm d}t_{\rm h},{\rm d}t_{\rm b})$ in our simulations, where ${\rm d}t_{\rm h}=\frac{{\rm d}x}{c_{\rm s}+|v|}$; 
${\rm d}x$ is the grid resolution, $c_s$ is the local sound speed, $v$ is the fluid velocity, and the time scale 
for burning is ${\rm d}t_{\rm b}$, which is determined by both the energy generation rate and the rate of change 
of the abundances.

\subsection{Homographic Expansion}

As we have shown, grid resolutions of $10^8\,\cm$ are needed to fully resolve nuclear burning 
in our model.  However, the star can have a radius of up to several $10^{14}\,\cm$.  This large dynamical range (10$^6$) 
makes it impractical to simulate the entire star at once
while fully resolving all relevant physical processes.  When the shock launches from the center of the star,
the shock's traveling time scale is about a few days, which is much shorter than the Kelvin--Helmholtz time
scale of the stars, about several million years.  We can assume that when the shock propagates inside the star, the 
stellar evolution of the outer envelope is frozen.  This allows us to trace the shock propagation without considering 
the overall stellar evolution.  Hence, we instead begin our simulations with a coordinate mesh that encloses just 
the core of the star with zones that are fine enough to resolve explosive burning.  We then halt the simulation as 
the SN shock approaches the grid boundaries, uniformly expand the simulation domain, and then restart the 
calculation.  In each expansion we retain the same number of grids  (see Figure \ref{expand}).  Although 
the resolution decreases after each expansion, it does not affect the results at later times because burning 
is complete before the first expansion and emergent fluid instabilities are well resolved in later expansions.  
These uniform expansions are repeated until the fluid instabilities cease to evolve.  There might be some
possible sound waves generated from boundaries under such a setup.  However, the normal SN shocks have 
a much higher mach number---above 10---while traveling inside the star.  The sound waves could not contaminate 
the burning/fluid instabilities domains before the shock reaches the boundary of the simulation box.   

Most stellar explosion problems need to deal with a large dynamic scale such as the case discussed here.  
It is computationally inefficient to simulate the entire star with a sufficient resolution.  Because the 
time scale of the explosion is much shorter than the dynamic time of stars, we can follow the evolution of the shock 
by starting from the center of the star and tracing it until the shock breaks out of the stellar surface. 
The utility of homographic expansion is also available in \CASTRO.

\begin{figure}[h]
\begin{center} 
\includegraphics[scale=0.6]{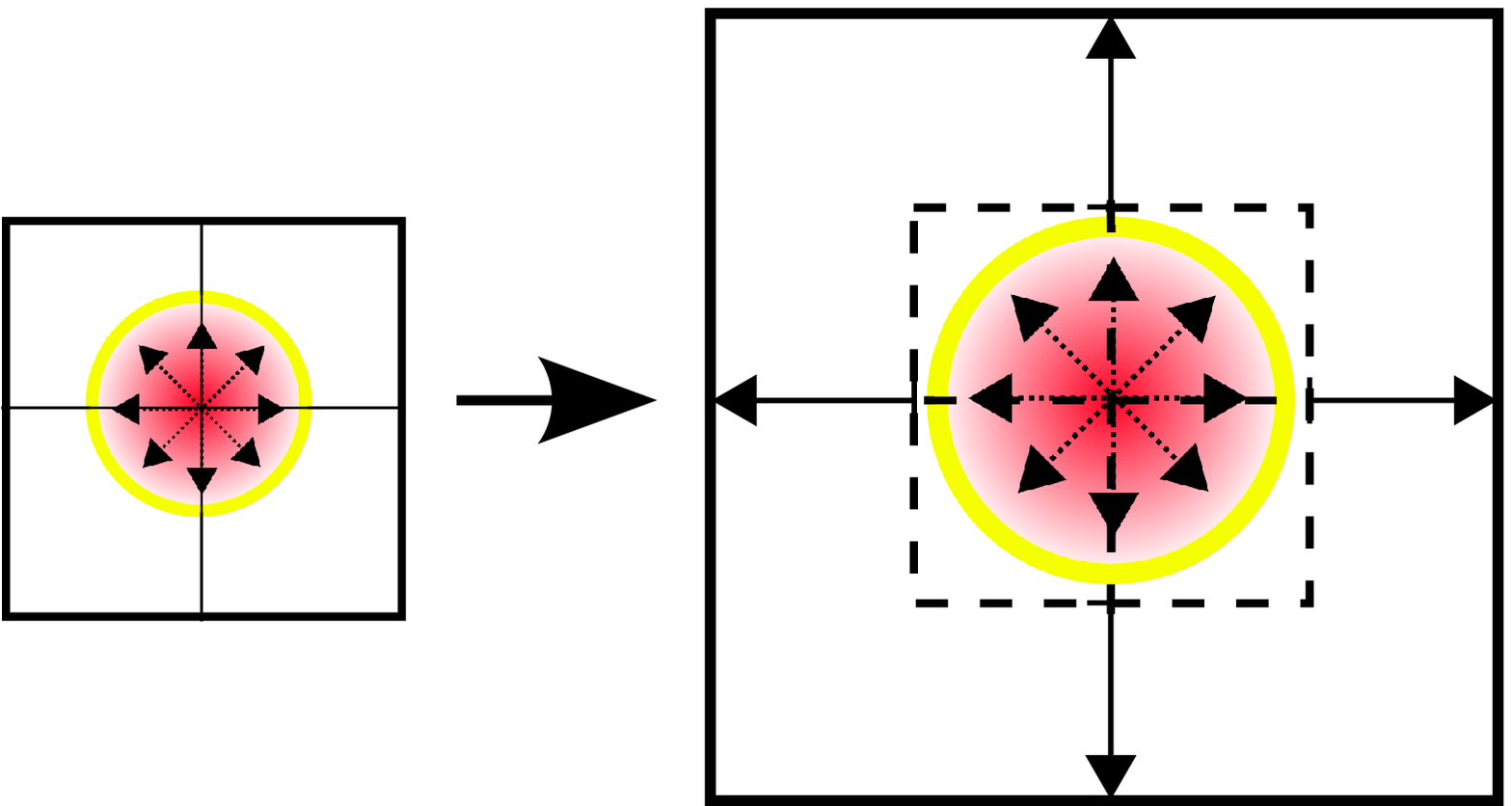}
\caption{Homographic expansion:  In both panels, the yellow circle is the SN shock and the red 
region is the ejecta.  The simulations begin with just the inner part of the star and a higher resolution 
(left panel) for capturing fluid instabilities and burning.  After the explosion occurs, we follow the shock 
until it reaches the boundary of the simulation box.  We then expand the simulation domain, 
mapping the final state of the previous calculation onto the new mesh with a new ambient 
medium that is taken from the initial profile.    \label{expand}}
\end{center}
\end{figure}

\section{Conclusion}
\lSect{conclusions}

Multidimensional stellar evolution and supernova simulations are numerically challenging because 
multiple physical processes (hydrodynamics, gravity, burning) occur on many scales in space and 
time.  For computational efficiency, 1D stellar models are often used as initial conditions in 2D and 
3D calculations.  Mapping 1D profiles onto multidimensional grids can introduce serious numerical 
artifacts, one of the most severe of which is the violation of conservation of physical quantities.  We 
have developed a new mapping algorithm that guarantees that conserved quantities are preserved 
at any resolution and it reproduces the most important features in the original profiles.  Our method is 
practical for 1D and 2D calculations, and we plan to develop integral methods (an explicit integral 
approach instead of using volume subsampling) that are numerically tractable in 3D.   

Multidimensional models give insight on fluid instabilities in supernova explosions that break the 
spherical symmetry of stars and mix their interiors.  These instabilities originate from perturbations in 
the star prior to the explosion.  Until now, these perturbations have been randomly seeded in 2D and 
3D models with little or no physical basis.  We present a new approach to seeding supernova models 
with physically realistic velocity perturbations like those found in the turbulent convective zones of 
massive stars.  We find that the initial spectrum of the perturbations tends to be smeared out as they 
become nonlinear.  Our approach can be applied to other multidimensional simulations of stellar explosions, 
especially those whose final outcomes are sensitive to the form of the initial perturbation; or the simulations of 
short duration, in which perturbations may not become fully nonlinear.  

Finally, we provide possible approaches to obtain the proper resolution for simulations that include 
both hydrodynamics and nuclear burning.  Because the burning changes both the internal energy 
and composition of the fluid, we determine the physical scale for resolving burning with resolution 
tests and proper time steps by considering both hydro and burning.  We apply a homographic 
expansion to bypass the numerical difficulties associated with the large range of dynamical scales in 
our problem.  The algorithms we present can be applied to other multidimensional simulations in addition to
stellar explosions in both astrophysics and cosmology.  

\section*{Acknowledgments}
The authors thank anonymous referees  for reviewing this manuscript and providing 
many insightful comments,  the members of the CCSE at LBNL for help with \CASTRO{}, 
and Hank Childs for assistance with {\it VISIT}.  We also thank Volker Bromm, Dan Kasen, 
Lars Bildsten, John Bell, Adam Burrows, and Stan Woosley for many useful discussions.  K.C. 
was supported by the IAU-Gruber Fellowship, Stanwood Johnston Fellowship, and 
KITP Graduate Fellowship. A.H.was supported by a future fellowship from the Australian 
Research Council (ARC FT 120100363). All numerical simulations were performed with allocations
from the University of Minnesota Supercomputing Institute and the National Energy Research Scientific
Computing Center.  This work has been supported by the DOE grants; DE-SC0010676, 
DE-AC02-05CH11231, DE-GF02-87ER40328, DE-FC02-09ER41618 and by the NSF grants;
 AST-1109394, and PHY02-16783.










\end{document}